\documentclass[useAMS,usegraphicx,usenatbib]{mn2e}
\usepackage{amsfonts,amssymb,amsmath}
\usepackage{xcolor}
\usepackage{graphicx}
\usepackage{soul}
\usepackage{epstopdf}
\usepackage{multirow}
\usepackage{multicol}
\newcommand{\hmpc}{h^{-1}{\rm Mpc}}

\newcommand{\msun}{{\rm M}_\odot}
\newcommand{\mpc}{{\rm Mpc}}
\newcommand{\kpc}{{\rm kpc}}
\newcommand{\kms}{{\rm ~km~s^{-1}}}
\newcommand{\magsec}{{\rm ~mag~arcsec^{-2}}}

\title[Compact Groups analysis using weak lensing]{Compact Groups analysis using weak gravitational lensing}
\author[Chalela et al.]{Mart\'in Chalela$^{1,2}$\thanks{E-mail: mchalela@oac.unc.edu.ar},
Elizabeth Johana Gonzalez $^{1,2,3}$,
\newauthor % starts a new line in the
Diego Garcia Lambas$^{1,3}$,
Gael Fo\"{e}x $^{4}$\\
$^{1}$ Instituto de Astronom\'{\i}a Te\'orica y Experimental, (IATE-CONICET),
 Laprida 854, X5000BGR, C\'ordoba, Argentina.\\
$^{2}$ Facultad de Matem\'atica, Astronom\'ia y F\'isica, FAMAF, Universidad Nacional de C\'ordoba, X5000BGR, C\'ordoba, Argentina.\\
$^{3}$ Observatorio Astron\'omico de C\'ordoba, Universidad Nacional de C\'ordoba, Laprida 854, X5000BGR, C\'ordoba, Argentina.\\
$^{4}$ Max Planck Institute for Extraterrestrial Physics, Giessenbachstrasse, 85748 Garching}

\begin{document}

%\date{Accepted 1988 December 15. Received 1988 December 14; in original form 1988 October 11}

\pagerange{\pageref{firstpage}--\pageref{lastpage}} \pubyear{2015}

\maketitle

\label{firstpage}

\begin{abstract}

We present a weak lensing analysis of a sample of SDSS Compact Groups (CGs). Using the measured radial density contrast profile, we derive the average masses under the assumption of spherical symmetry, obtaining a velocity dispersion for the Singular Isothermal Spherical model, $\sigma_V = 270 \pm 40 \kms$, and for the NFW model, $R_{200}=0.53\pm0.10\,h_{70}^{-1}\,$Mpc. We test three different definitions of CGs centres to identify which best traces the true dark matter halo centre, concluding that a luminosity weighted centre is the most suitable choice. We also study the lensing signal dependence on CGs physical radius, group surface brightness, and morphological mixing. We find that groups with more concentrated galaxy members show steeper mass profiles and larger velocity dispersions. We argue that both, a possible lower fraction of interloper and a true steeper profile, could be playing a role in this effect.  Straightforward velocity dispersion estimates from member spectroscopy yields $\sigma_V \approx 230 \kms$ in agreement with our lensing results.

\end{abstract}


%====================================================================================
\section{INTRODUCTION}
%General introduction about compact groups.
%
The largest concentrations of mass and visible matter in the Universe reside in galaxy clusters. However, a significant fraction of galaxies are located in groups of different mass and morphology content \citep{Karachentsev2005}. Studying the physical properties of these systems is of prime importance to understand galaxy formation and evolution.

Compact groups of galaxies (CGs) are a special class of galaxy systems, containing generally 4 to 6 members within a region of just a few galaxy radii, and with a low radial velocity dispersions \citep[$\sim 200\kms$, e.g.][]{McConnachie2009}. This particular combination implies that CGs have short crossing times ($\sim 0.2$ Gyr), providing an ideal scenario to study galaxy merging and the impact of enviroment on galaxy evolution. However, the effects of such an extreme environment and the short time-scales in which these systems would collapse are not completely understood, setting an ongoing debate about the nature of these systems. Numerical simulations have shown that member galaxies can eventually merge and so groups may disappears \citep{Barnes1985, Barnes1989, Mamon1987} in a time scale comparable to the observed crossing times \citep{Hickson1992}. Other simulations present an alternative picture, where CGs lifetime is much longer than the crossing time \citep{Governato1991, Athanassoula1997} which would explain the relatively high number density of these systems in the observations. Nevertheless, there is a strong debate regarding the genuineness of these systems, since it has been suggested that most of them could be spurious line-of-sight alignments rather than truly bound systems \citep{Mamon1986}.

In a widely accepted scenario, CGs are gravitationally bound, but unstable systems. The X-ray observations showing great emission from the hot intragroup gas \citep{Ponman1996}, suggest that strong interactions between member galaxies could have provided a significant intragroup medium. Orbital decay due to dynamical friction should strip away galaxies from their haloes resulting in eventual mergers in short timescales, leading to a morphological evolution. Therefore, the fraction of early-type galaxies would pinpoint the evolutionary state of the groups as a whole. Although group members can merge, CGs may increase their number of members by acquiring them from the surroundings, extending their lifetime \citep{Diaferio1994}. Many studies support this scenario showing that most of these galaxy systems reside within larger structures such as loose groups and rich clusters \citep[e.g.][]{Rood1994, deCarvalho2005, Mendel2011}.

Hickson CGs \citep[hereafter HCG, ][]{Hickson1982} sample has been widely analysed providing several studies of these systems at low redshift ($z \sim 0.03$). High mass-to-light ratio determinations of $50h \Upsilon_{\odot}$ and typical line-of-sight velocity dispersions of $200\kms$ \citep{Hickson1992}, suggests the presence of substantial amounts of dark matter. Furthermore, a recent study by \citet{Pompei2012}, based on spectroscopically confirmed CGs at higher redshift ($z \sim 0.12$), reports remarkably higher average values of $M/L_{B} = 190\Upsilon_{\odot}$ and $\sigma_{LOS} = 273 \kms$. The authors suggest these high values could be due to the proximity of large-scale structures, which may affect mass estimates. Despite differences with other authors, these results are consistent with predictions of the hierarchical model of structure formation. Results from hydrodynamical and N-body simulations show that individual dark matter haloes of CGs members merge first, creating a common massive halo that dominates galaxy dynamics \citep{Barnes1984, Bode1993}.

Until now CGs' masses have been determined through a dynamical approach, either by measuring velocity dispersions or through X-ray observations. \citet{Ponman1996} showed that these systems slightly deviate from the known relation $L_{X}-T$ for clusters (being fainter than the predicted one) but are still consistent with the $L_{X}-\sigma_{LOS}$ relation. Gravitational lensing provides an alternative approach to measure the mass of galaxy systems. \citet{MendesdeOliveira1994} analysed the possibility that a CG could act as a lensing system. Based on the HCGs the authors quantified the lensing efficiency, concluding that they would be too weak to be detected as a lens since this sample is quite nearby. However, their calculations show that CGs at higher redshifts ($z \sim 0.1$), such as those available in modern catalogues, could produce a detectable lensing signal.

Weak lensing techniques have been applied almost exclusively to clusters of galaxies providing precise determinations consistent with values derived from dynamical analysis and X-ray observations \citep{Hoekstra1998, Fischer1999, Clowe2000}. In recent years, several studies have analysed the lensing effects produced by groups of galaxies \citep[e.g.][]{George2012, Spinelli2012, Foex2013, Foex2014}, nevertheless none of them have focused on CGs. In order to apply weak lensing techniques to low mass galaxy systems, such as groups with masses $\sim 10^{13} M_{\odot} $, stacking techniques have shown to be a powerful tool to increase the signal-to-noise ratio and thus, suitable to derive groups statistical properties \citep[e.g.][]{Rykoff2008, Leauthaud2010, Foex2014}.

In this work we present the first statistical weak lensing analysis of a sample of CGs using stacking techniques. Our systems were extracted from the catalogue of CGs of \citet{McConnachie2009}. Images for the analysis were obtained from Sloan Digital Sky Survey data \citep{York2000}. This survey has the largest imaging coverage available at present, providing a statistically significant data base suitable for stacking techniques. These data have been successfully used in previous weak lensing studies to analyse the density profile and determine total masses of galaxies and galaxy systems \citep[e.g., ][]{Mandelbaum2006, Sheldon2009, Clampitt2016, Eli2016}. From our lensing analysis, we derive the average mass under the assumption of spherical symmetry. We probe three different definitions of CGs centre to identify which one best traces the dark matter halo. Furthermore, we compare our results with dynamical estimates and we analyse the observed lensing signal according to various CGs properties. The paper is organized as follows. In section \ref{sec:sample} we describe the selection of groups used throughout the study. In section \ref{sec:methodology} we briefly describe the weak lensing analysis, as this was extensively discussed in previous works, along with the formalism of miscentred density profiles. In section \ref{sec:results} we present the obtained mass and finally, in section \ref{sec:discussion} we summarise our results and compare them with other studies. We adopt, when necessary a standard cosmological model $H_{0} = 70\kms\mpc^{-1}$, $\Omega_{m} = 0.3$ and $\Omega_{\Lambda} = 0.7$.

%====================================================================================
\section{COMPACT GROUPS: SAMPLE DESCRIPTION AND SOURCE GALAXIES}		\label{sec:sample}
%Description of the CFHT and SDSS samples.
\subsection{McConachie Compact Groups}

There are several catalogues of compact groups in the literature. In general, the identification of these data sets follow Hickson's original selection criteria, or variations in order to identify similar systems. Some are based on spectroscopic information like \citet{Barton1996} and \citet{Allam2000}, while others follow photometric criteria such as \citet{Hickson1982}, \citet{Prandoni1994}, \citet{Iovino2002} and \citet{McConnachie2009}. In order to statistically increase the lensing signal, the weak lensing analysis requires stacking of a large number of CGs. We have used \citet{McConnachie2009} catalogue, which comprises the largest CGs sample available at present. This catalogue is based on photometric data from the sixth data release of the Sloan Digital Sky Survey \citep[SDSS-DR6, ][]{DR6}. CGs were identified by applying \citet{Hickson1982} criteria, where member  galaxies satisfy:
\begin{enumerate}
\item $N (\Delta m = 3) \geq 4$;
\item $\theta_N 	\geq 3 \theta_G$;
\item $\mu \leq 26.0\,mag\,arcsec^{-2}$
\end{enumerate}
$N (\Delta m = 3)$ is the number of member galaxies within 3 magnitudes of the brightest galaxy, $\theta_G$ is the angular diameter of the smallest circle that enclose the centres of these galaxies, $\theta_N$ is the angular diameter of the largest concentric circle with no additional galaxy in this magnitude range or brighter, and $\mu$ is the effective surface brightness of member galaxies (where the total flux is averaged over the circle of angular diameter $\theta_G$).

These criteria were applied in two ranges of limiting magnitude resulting in two datasets, Catalogue A and Catalogue B. Catalogue A includes 2297 CGs identified from galaxies with $r$ magnitude in the range $14.5 \leq r \leq 18.0$. Catalogue B contains 74791 CGs with member galaxies in a wider magnitude range $14.5 \leq r \leq 21.0$. An individual visual inspection of all groups in Catalogue A was carried out minimizing the contamination of the sample due to photometric errors in the automatic SDSS pipelines. This procedure was not applied to Catalogue B given the large number of objects, with an estimated contamination by false sources of about $14\%$. Both catalogues provide detailed information about CGs and their member galaxies such as group surface brightness, radius and number of members, as well as each galaxy $r$ and $g$ magnitude, and spectroscopic redshift (when available). Given that the Hickson criterion relies only on photometric information, not all CG members may have spectroscopic data.

%=================================================================
\begin{figure*}
\centering
\includegraphics[scale=0.8]{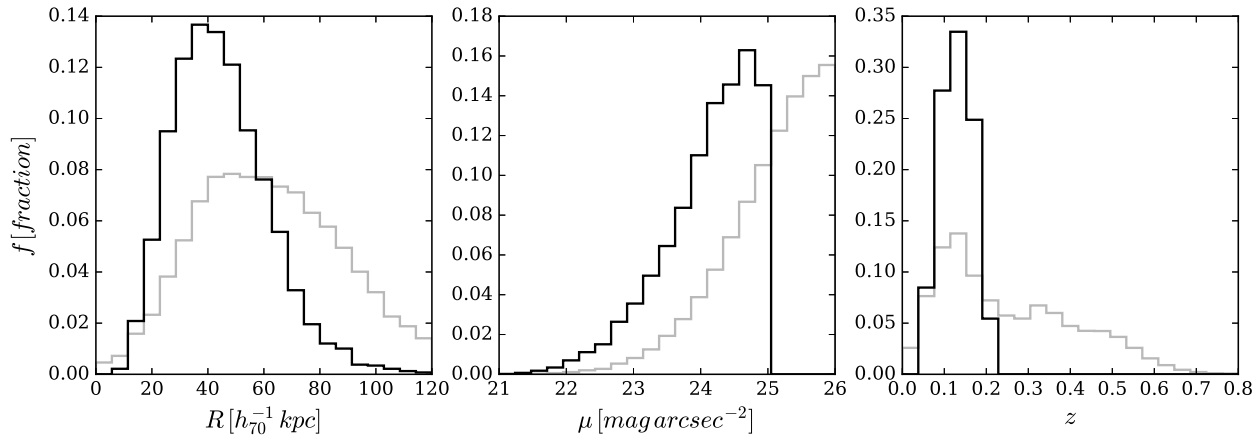}
\caption{Normalized distributions of parameters of the analysed CGs (black line) and catalogue B (gray line). From left to right: physical radius $R$, surface brightness $\mu$ and redshift $z$.}
\label{fig:sample}
\end{figure*}
%=================================================================
\subsection{Final sample and image data}
For statistical reasons we extracted our sample from Catalogue B. Redshifts of all galaxies in this catalogue were updated with information from SDSS Data Release 12, and we recalculated CGs redshifts as the mean value of the group members. The redshift distribution of the updated catalogue B peaks at $z \approx 0.1$ extending up to $z \approx 0.6$. 

Given that the lensing efficiency depends on the lens distance and considering that the redshift distribution peaks at $z \sim 0.1$, we discard groups with $z< 0.06$ which contribute little weight. We also discard systems with $z > 0.2$ since the density of background galaxies is insufficient to extract a reliable signal. We analyse only objects with $\mu \leq 25 \magsec$, where $\mu$ is defined as the $r$-band surface brightness. This cut is made to increase the fraction of CGs without interlopers in the sample; members of brighter groups are more probable to be part of a real bound system and not a visual alignment in the sky. According to \citet{McConnachie2008} the sample purity improves from about $30\%$, for CGs with $\mu \leq 26 \magsec$, to $43\%$, for groups with $\mu \leq 25 \magsec$. 

The final sample consists of 6257 CGs. In Figure \ref{fig:sample} we show the distribution of CGs properties of catalogue B and our final sample. It can be noticed that with the mentioned cuts, we exclude the more extended ($R \gtrsim 80 \,h^{-1}_{70}\, \kpc$) CGs.

Image data was obtained from the SDSS. This survey provides the largest photometric and spectroscopic public database available at present. It was constructed using a 2.5 m telescope at Apache Point Observatory in New Mexico. The tenth data release \citep[SDSS-DR10, ][]{Ahn2014} covers 14555 square degrees of sky imaged in five bands ($u, g, r, i$ and $z$) and has a limiting magnitude $r = 22.2$. For the lensing analysis we use images in $r$ and $i$ band, obtained from DR10 as it includes all prior SDSS imaging data. This allows us to select the frame with the best seeing conditions in the field of a given CG. Each SDSS image is $9.8' \times 13.5'$, corresponding to $1489 \times 2048$ pixels, with a pixel size of $0.396''$. The average seeing is about $1''$ in the $i$-band.

\subsection{Photometry, source classification and shape measurements}

In this subsection we describe the details regarding detection, classification and shape measurements of background galaxies. The implemented pipeline has been successfully applied to SDSS data in order to estimate total masses of galaxy systems \citep{Eli2016}.

We conduct a search of frames in order to analyse the most adequate images for  our lensing analysis. Thus, for each CG we sequentially search and retrieve the best centred $i$-band frames within 50 pixels from the borders and select the first frame in the search with seeing lower than $0.9''$. If no frame satisfies this seeing condition, we choose that with the lowest seeing, up to $1.3''$. CGs in frames not satisfying seeing values $<1.3''$ are discarded. This results in 5568 CGs suitable for the analysis (i.e. $\sim$90\% of the selected 6257 systems). After the $i$-band frame is selected we also retrieve the corresponding $r$-band frame. Notice that given the low lensing signal expected at large radii from the lens centre, it is not necesary to use a frame mosaic, but rather use a single frame for each system.

To perform the detection and photometry of the sources we implement \texttt{SExtractor} \citep{Bertin1996} as described in \citet{Eli2015}, in a two-pass mode. The first run is made with a detection level of $5\sigma$ above the background to detect bright objects and estimate the seeing. A second run is made with a detection level of $1.5\sigma$ in dual mode to detect objects on the $i$ frame, while photometric parameters are measured on both $i$ and $r$-band frames.

\begin{figure}
\centering
\includegraphics[scale=0.4]{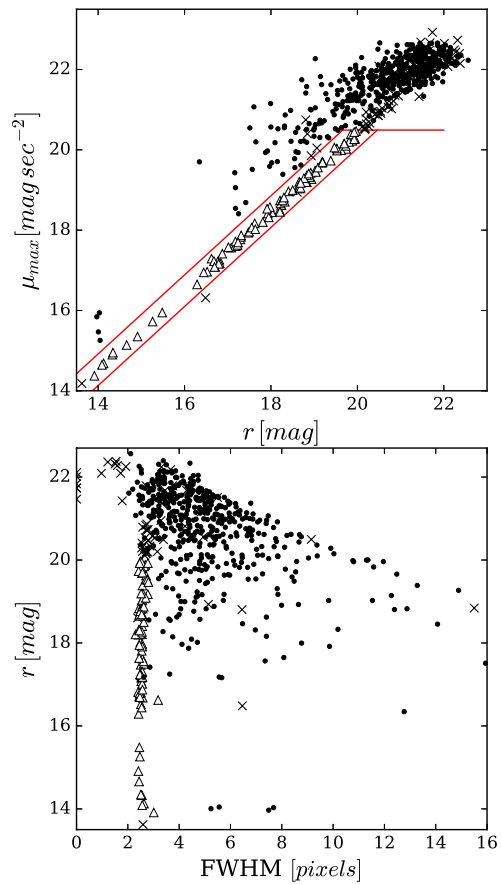}
\caption{Source classification in a frame with seeing $\approx 1.0''$. Stars and galaxies are represented with triangles and dots, respectively; false detections are marked with crosses. In the top panel we show the magnitud-peak surface brightness scatter plots. Stars are located in the enclosed region (see text) limited to a maximum $\mu_{max}$ value where galaxies start to overlap the star sequence. Sources at the fainter side of this region are considered as false detections. In the bottom panel we show the FWHM-magnitude scatter plot.}
\label{fig:classification}
\end{figure}

Sources are classified in stars, galaxies and false detections according to their full-width (FWHM), stellarity index and position in the magnitude-peak surface brightness ($\mu_{max}$) plot, where these parameters are obtained from \texttt{SExtractor} output. In Figure \ref{fig:classification} we show an example of the source classification for a single frame whith seeing $=1.0''$. Objects that are more sharply peaked than the point spread function (PSF), thus with FWHM $<$ seeing $- 0.5$ pixel, and with \texttt{SExtractor} FLAG parameter $>$ 4, are considered as false detections. As the light distribution of a point source scales with magnitude, objects on the magnitude-$\mu_{max}$ line $\pm$ 0.4 mag and FWHM $<$ seeing + 0.8 pixel are considered as stars. The rest of the sources with stellarity index $<$ 0.8 are classified as galaxies.

\begin{figure}
\centering
\includegraphics[scale=0.4]{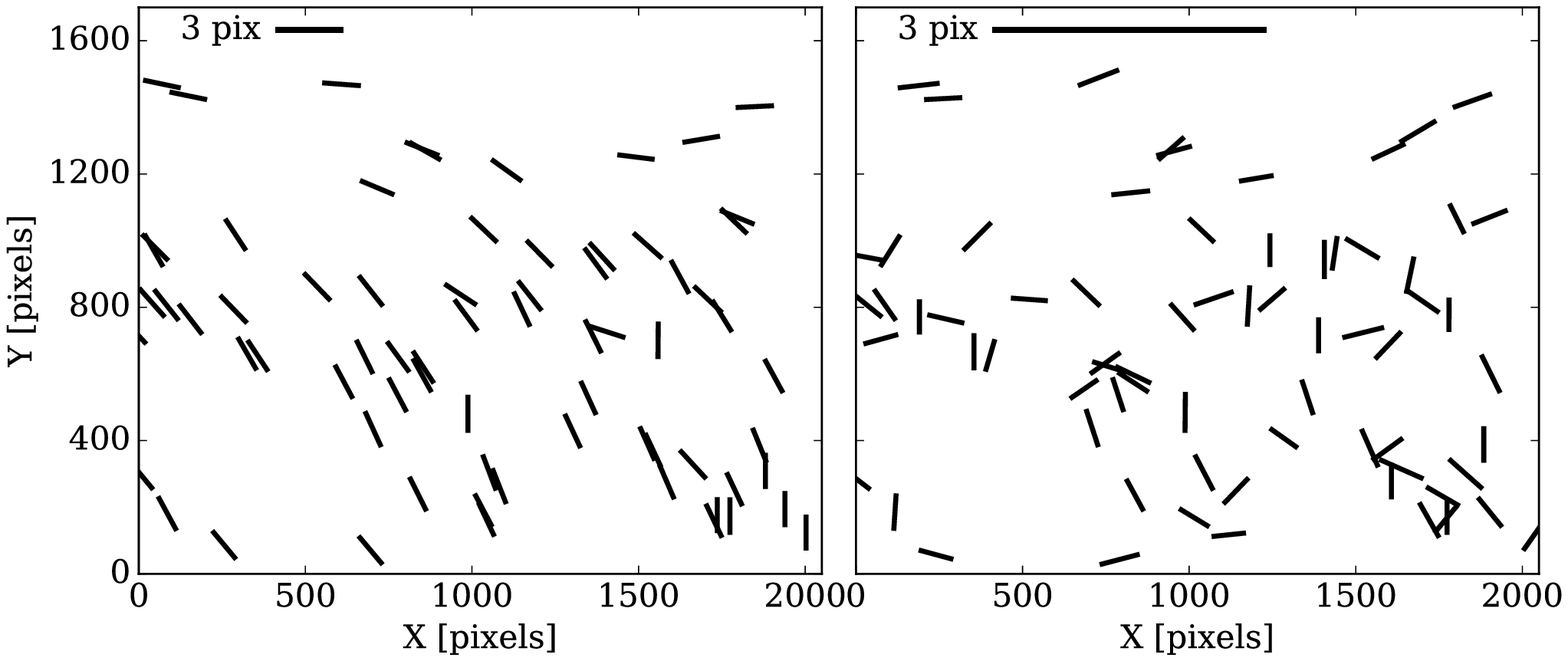}       
\includegraphics[scale=0.4]{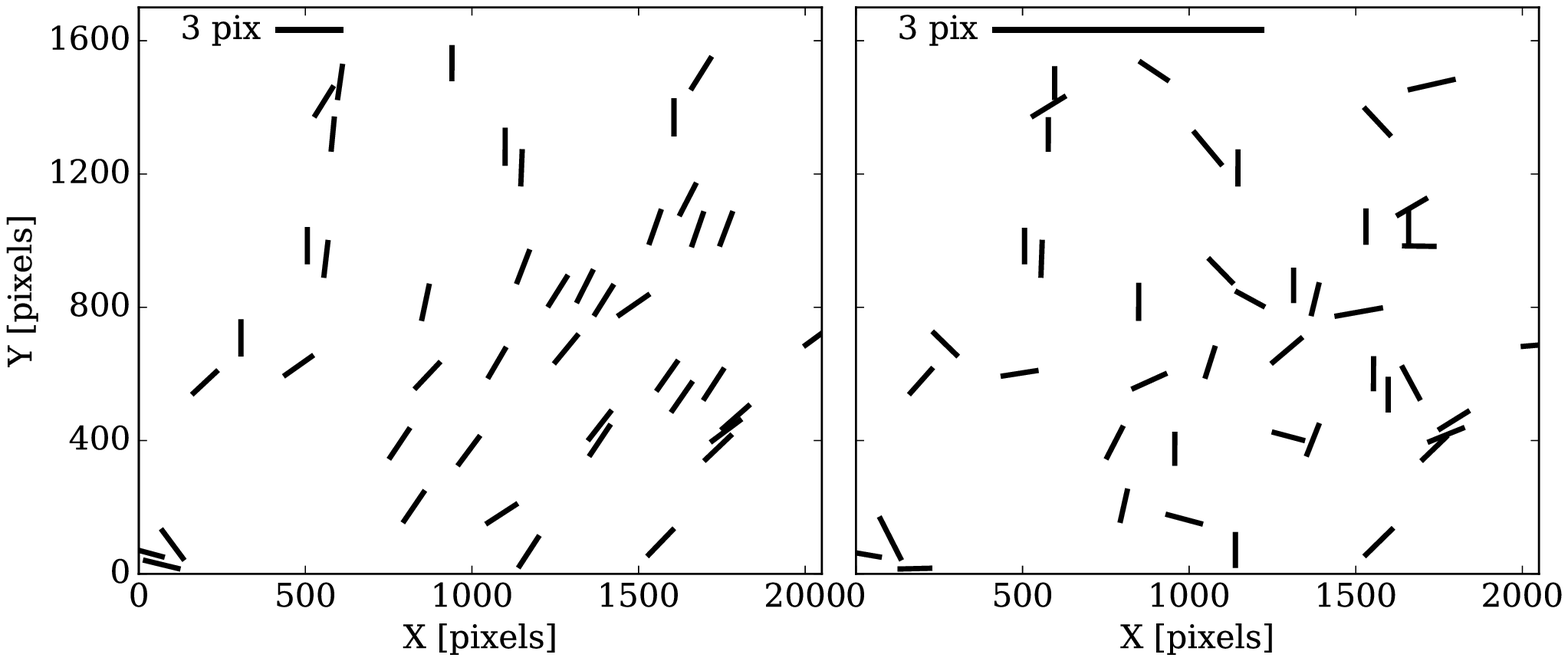}
\caption{PSF correction applied to stars of two frames: semi-major axes before (left panels) and after (right panels) the deconvolution. Notice that after taking into account the PSF correction, semi-major axes orientations are randomly distributed and with significantly smaller moduli.}
\label{fig:PSF}
\end{figure}

For the shape measurements we use \texttt{Im2Shape} \citep{Bridle2002} which computes the shape parameters modelling the object as sum of gaussians convolved with a PSF, also modeled as a sum gaussian. For simplicity, we modeled the sources and the PSF using only one gaussian. The PSF map across the image is estimated from the shape of stars, since they are intrinsically point-like objects. We only used objects with a measured ellipticity smaller than 0.2 to remove most of the remaining false detections and faint galaxies. Looking at the five nearest neighbours of each star, we also removed those that differ by more than $2\sigma$ from the local average shape. Finally, the local PSF at each galaxy position is linearly interpolated by averaging the shapes of the five nearest stars. Once the PSF is determined, we run \texttt{Im2Shape} on galaxies to meassure their intrisic shape parameters. In order to test our PSF treatment, we apply the PSF correction on stars to check that it can recover point-like shapes. In Figure \ref{fig:PSF}, we show the major semi-axis distribution of stars for two frames, before and after taking into account the PSF in the shape measurement. After the PSF correction, the major semi-axis sizes are considerably smaller and the orientation is randomly distributed, consistent with point-like sources.

\begin{figure}
\centering
\includegraphics[scale=0.4]{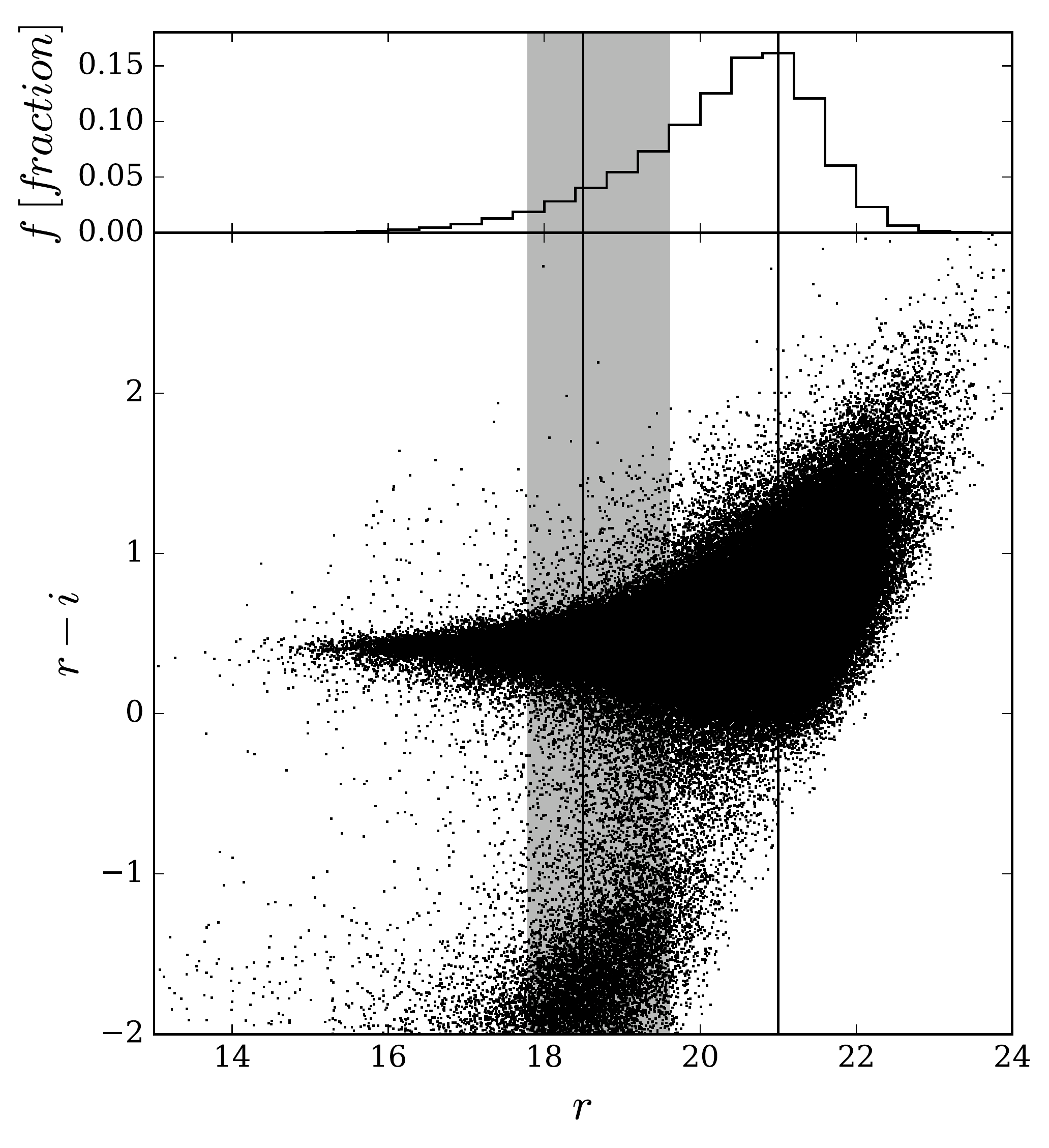}
\caption{Color-magnitude diagram (bottom) and normalized magnitude distribution (top) of sources classified as galaxies in the CGs fields. The vertical lines indicate the magnitude cuts used for the selection of background galaxies. The shaded region spans the entire $m_P$ range and the inner line indicates the mean value, $\langle m_P \rangle$. The solid line at $r=21$ indicates the faint limit cut.}
\label{fig:color_mag}
\end{figure}

To perform the lensing analysis, background galaxies are selected as those with $r$ magnitudes between $m_P$ and 21 mag. $m_P$ is defined as the faintest magnitude at which the probability that a galaxy is behind the group is higher than 0.7. This value is computed according to the redshift of each CG using a catalogue of photometric redshifts \citep[see][for details about $m_P$ estimation]{Eli2015}. Discarding galaxies fainter than 21 mag ensures that we are not taking into account faint galaxies with high uncertainties in their shape measurements. We also restrict the selection to those objects with a good pixel sampling by using only galaxies with FWHM $>$ 5 pixels. In Figure \ref{fig:color_mag} we show the color-magnitude diagram of all selected galaxies with the photometric cuts used for the background galaxy selection. The average number of backgorund galaxies obtained is 60 per frame, which corresponds to a density of $\sim 0.46$ galaxies/arcmin$^2$, making a total of $\sim 2600$ galaxies/arcmin$^2$ for the catalogue used in the stacking analysis.

%=================================================================
%=================================================================
%=================================================================

%=================================================================

%====================================================================================
\section{WEAK LENSING METHODOLOGY}		\label{sec:methodology}
\subsection{Stacking technique}
%Description of our weak lensing analysis: pipeline and stacking.
We briefly describe the lensing analysis and the stacking technique as these were described in detail in \citet{Eli2015, Eli2016}. Gravitational lensing effects are characterized by an isotropic stretching called convergence, $\kappa$, and an anisotropic distortion called shear, $\gamma$. Using the second derivative of the projected gravitational potential to express the shear and convergence, one can show that for a lens with a circular-symmetric projected mass distribution, the tangential component of $\gamma$ is related to the convergence through \citep{Bartelmann1995}:
\begin{equation}
\gamma_T(r) = \bar{\kappa}(<r) - \bar{\kappa}(r)
\end{equation}
where $\bar{\kappa}(<r) $ and $\bar{\kappa}(r)$ are the convergence averaged over the disk
and circle of radius $r$, respectively. On the other hand, the cross component of the shear, $\gamma_{\times}$, defined as the component tilted at $\pi$/4 relative to the tangential component, should be exactly zero.

Since the convergence is defined as the surface mass density $\Sigma(r)$ normalized by the critical density $\Sigma_{crit}$ , we can rewrite the previous equation defining the density contrast, $\Delta\tilde{\Sigma}$, which is redshift-independent:
\begin{equation}
\label{Dsigma}
\tilde{\gamma}_{T}(r) \times \Sigma_{crit} = \bar{\Sigma}(<r) - \bar{\Sigma}(r) \equiv \Delta\tilde{\Sigma}(r)
\end{equation}

%Then we can estimate the average tangential shear, $\tilde{\gamma}_{T}$, induced in a population of background galaxies by the presence of a foreground CG and calculate the density contrast.

The tangential shear component is directly estimated as $\tilde{\gamma}_{T} = \langle e_{T} \rangle$, where the tangential ellipticity of background galaxies is averaged over annular bins. The averaged cross ellipticity component, in turn, should be zero and corresponds to the cross shear component.

For the composite lens, the density contrast is obtained as the weighted average of the tangential ellipticity of background galaxies:
\begin{equation}
\langle \Delta \tilde{\Sigma}(r) \rangle = \frac{\sum_{j=1}^{N_{Lens}} \sum_{i=1}^{N_{Sources,j}} \omega_{ij} \times e_{T,ij} \times \Sigma_{crit,j}}{\sum_{j=1}^{N_{Lens}} \sum_{i=1}^{N_{Sources,j}} \omega_{ij}}
\end{equation}
where $\omega_{ij}$ is the associated weight of each background galaxy as described in \citet{Eli2016}. $N_{Lens}$ is the number of lensing systems and $N_{Sources, j}$ the number of background galaxies located at a distance $r \pm \delta r$ from the $j$th lens. $\Sigma_{crit,j}$ is the critical density for all the sources of the lens $j$, defined as:
\begin{equation*}
\Sigma_{crit,j} = \dfrac{c^{2}}{4 \pi G} \dfrac{1}{\langle \beta_{j} \rangle D_{OL_{j}} }
\end{equation*}
Here $D_{OL_{j}}$ is the angular diameter distance from the observer to the $j$th lens, $G$ is the gravitational constant, $c$ is the light velocity and $\langle \beta_{j} \rangle$ is the geometrical factor defined as the average ratio between the angular diameter distance from the galaxy source $i$ to the lensing system $j$ and the angular diameter distance between the observer and the source ($\langle \beta_{j} \rangle = \langle D_{LS_{j}}/D_{OS_{i}} \rangle_{i}$). Given the lack of redshift information for individual background galaxies, it is not possible to directly estimate the geometrical factor $\beta$. Therefore, we estimated this value using \citet{Coupon2009} catalogue of photometric redshifts. This catalogue is based on the public release Deep Field 1 of the Canada-France-Hawaii Telescope Legacy Survey, which is complete down to $m_r = 26$. We computed $\langle \beta_{j} \rangle$ after applying the same photometric cut used in the selection of background galaxies. This value is fairly insensitive to the detailed redshift distribution, as long as the mean redshift of background galaxies is considerably larger than the lens redshift \citep{libro}. This is the case of our sample, which has a mean redshift of 0.1, while the mean redshift of background galaxies is 0.32. We consider the contamination due to foreground galaxies by setting $\beta(z_{phot} < z_{lens}) = 0$, which outbalances the dilution of the shear signal by these unlensed galaxies. The average $\langle \beta_{j} \rangle$ value is $\approx0.50$.

The misidentification of faint group members as background galaxies weakens the lensing signal since they are not sheared. Although CGs have few members, numerical simulations suggests that fainter satellite galaxies could be surrounding the group. To overcome this problem, $\langle \Delta \tilde{\Sigma}(r) \rangle$ is multiplied by a factor $1+f_{cg}(r)$ following \citet{Hoekstra2007}, where $f_{cg}(r)$ is the fraction of group members that remains in the catalogue of background galaxies. To estimate $f_{cg}(r)$ we fit a $1/r$ profile to the galaxy excess relative to the background level and we correct the measured shear according to the distance to the lensing system centre.

The statistical uncertainties associated with the estimator $\langle \Delta \tilde{\Sigma}(r) \rangle$ are computed taking into account the noise due to the galaxies' intrinsic ellipticity:
\begin{equation} 
\label{eq:err}
\sigma^{2}_{\Delta \tilde{\Sigma}}(r)=\dfrac{\sum_{j=1}^{N_{Lens}} \sum_{i=1}^{N_{Sources,j}}( \omega_{ij} \times \sigma_{\epsilon} \times \Sigma_{crit,j})^{2}}{(\sum_{j=1}^{N_{Lens}} \sum_{i=1}^{N_{Sources,j}}  \omega_{ij})^{2}}
\end{equation}
where $\sigma_{\epsilon}$ is the dispersion of the intrinsic ellipticity distribution. We adopt $\sigma_\epsilon=0.32$ according to the value considered by \citet{Clampitt2016} for a sample of background galaxies measured using SDSS data image. These quantities allow us to compute the total signal-to-noise ratio (S/N) as follows:
\begin{equation} 
\label{SN}
\left(\frac{S}{N}\right)^{2}=\sum_{i} \dfrac{ \langle \Delta \tilde{\Sigma}(r_{i})\rangle ^{2}}{\sigma^{2}_{\Delta \tilde{\Sigma}}(r_{i})}
\end{equation}
where the sum runs over all the bins used to fit the profile.

\subsection{Miscentred density contrast profile}
%Description of the miscentering problem

\citet{McConnachie2009} defines the centre of a CG as the centre of the smallest circle that contains the geometrical centre of its member galaxies. This position could be displaced from the true dark matter halo centre, leading to a flattening of the average density contrast profile and a mass underestimation.

If $r_{s}$ is the projected offset in the lens plane, the azimuthally averaged $\Sigma(r)$ profile is given by the convolution \citep{Yang2006}:
\begin{equation}
\Sigma(r|r_{s}) = \dfrac{1}{2\pi} \int_{0}^{2\pi} \Sigma(\sqrt{r^{2}+r_{s}^{2}+2r\,r_{s}\cos{\theta}}) d\theta.
\end{equation}
Since the actual offsets are not known, we adopt \citet{Johnston2007} approximation where a 2D gaussian distribution describes this miscentering:
\begin{equation}
P(r_{s}) = \dfrac{r_{s}}{\sigma_{s}^{2}}\exp{(-\dfrac{1}{2}(r_{s}/\sigma_{s})^{2})}
\end{equation}
where $\sigma_{s}$ is the width of the distribution. This value has been obtained in previous analysis of groups and cluster of galaxies, considering the BCG (\textit{Brightest  Cluster Galaxy}) of the system centre. \citep{George2012} reported $\sigma_s = 24.8 \pm 12 $\,kpc for  X-ray selected groups. On the other hand, other works estimate higher values ranging from $ 0.2 \hmpc$ to $ 0.42 \hmpc$, being higher for massive clusters \citep{Johnston2007,VanUitert2016}. The discrepancy between these results could rely on the sample properties, since X-ray selected groups may contain more relaxed systems. Taking into account the above considerations and the fact that CGs are much smaller than clusters, with typical radii of $\sim 40 \,h^{-1}_{70}\, \kpc$, we assume  $\sigma_{s} = 40 \,h^{-1}_{70}\, \kpc$. 

The resulting projected surface mass density for the sample can be written as
\begin{equation}
\Sigma_{s}(r) = \int_{0}^{\infty} P(r_{s})\Sigma(r|r_{s}) dr_{s}
\end{equation}
and $\Delta \Sigma_{s}(r)$ can then be calculated with \eqref{Dsigma} considering that:
\begin{equation*}
\bar{\Sigma}_{s}(<r) = \dfrac{2}{r^{2}} \int_{0}^{r} r' \Sigma_{s}(r') dr'.
\end{equation*}
The effect of this miscentring on $\Delta\Sigma(r)$ produces a suppression on the lensing signal at scales of the order of $\sigma_{s}$. On the outer region however, the signal remains almost unaffected.

\subsection{Fitting mass density profiles}
%------------------------------------------------------------------------------

Density contrast profile $\langle \Delta \tilde{\Sigma}(r_i)\rangle$ is computed using non-overlapping concentric logarithmic annuli to preserve the signal-to-noise ratio of the outer region, from $r_{in} = 50 \,h^{-1}_{70}\, \kpc$ up to $r_{out} \approx 900 \,h^{-1}_{70}\, \kpc$, where the signal weakens. We fit this profile using two models, the singular isothermal sphere (SIS) and the \citet[NFW,][]{Navarro1997} profile. The SIS profile describes a relaxed spherical distribution with a constant 1-D velocity dispersion, $\sigma_{V}$. In this model, the shear $\gamma (\theta)$ at an angular distance $\theta$ from the lens' centre, is directly related to $\sigma_{V}$ by the equation
\begin{equation}
\gamma (\theta) = \dfrac{\theta_{E}}{2 \theta}
\end{equation}
where $\theta_{E}$ is the critical Einstein radius defined as:
\begin{equation}
\theta_{E} = \dfrac{4 \pi \sigma_{V}^{2}}{c^{2}} \frac{1}{\langle \beta \rangle}
\end{equation}
From this model we can compute the characteristic mass $M_{200}\equiv\,M\,(\,<\,R_{200})$, defined as the mass within the radius that encloses a mean density 200 times the critical density of the universe, as in \citep{Leonard2010}:
\begin{equation}\label{eq:MSIS}
M_{200} =  \dfrac{2 \sigma_{V}^{3} }{\sqrt{50} G H(z)} 
\end{equation}
The NFW is a radial profile constructed by fitting the average halo density profile in cold dark matter numerical simulations. It depends on two parameters, $R_{200}$ and a dimensionless concentration parameter, $c_{200}$, as follows:
\begin{equation}
\rho(r) =  \dfrac{\rho_{crit} \delta_{c}}{(r/r_{s})(1+r/r_{s})^{2}} 
\end{equation}
where $r_{s}$ is the scale radius, $r_{s} = R_{200}/c_{200}$ and $\delta_{c}$ is the cha\-rac\-te\-ris\-tic overdensity of the halo,
\begin{equation}
\delta_{c} = \frac{200}{3} \dfrac{c_{200}^{3}}{\ln(1+c_{200})-c_{200}/(1+c_{200})}  
\end{equation}
In order to fit this profile, we use the gravitational lensing expressions formulated by \citet{Wright2000}. There is a well-known degeneracy between the two parameters $R_{200}$ and $c_{200}$ that can be broken by combining weak and strong lensing information. Since we lack of strong lensing information for CGs, we can estimate the concentration parameter with the relation $c_{200}(M_{200}, z)$, given by \citet{Duffy2011}, using the $M_{200}$ value obtained in the SIS fit and the average redshift of CGs weighted by their number of background galaxies. We use this aproximation considering that the derived NFW masses are not too sensitive to this parameter given the uncertainties in the shear profile. Once the concentration is estimated, we fit the NFW profile with only one free parameter, $R_{200}$, and calculate $M_{200}$.

We derived the parameters of each mass model performing a standard $\chi^{2}$ minimization:
\begin{equation}
\chi^{2} = \sum^{N}_{i} \dfrac{(\langle \Delta\tilde{\Sigma}(r_{i})  \rangle - \Delta\tilde{\Sigma}(r_{i},p))^{2}}{\sigma^{2}_{\Delta \tilde{\Sigma}}(r_{i})}
\end{equation}
where the sum runs over the $N$ radial bins of the profile and $p$ is the fitted parameter ($\bar{\sigma}_{V}$ in the case of the SIS profile, and $\bar{R}_{200}$ for the NFW model). Errors in the fitted parameters were computed according to the $\chi^{2}$ dispersion. The optimal bin steps were chosen to minimize $\chi^{2}$ values.

Other lensing studies consider the average density contrast profile taking into account the contribution from other neighboring mass concentrations by introducing another halo term \citep[e.g.,][]{Johnston2007,Leauthaud2010,Oguri2011}. In order to test our results derived up to $r_{out} = 900h_{70}^{-1}kpc$, we have also fitted the profiles within a significantly smaller radius ($r_{out} = 500h_{70}^{-1}kpc$). We find that the derived CGs density contrast profiles are in good agreement within uncertainties, showing the reliability of our results.

\subsection{Systematic errors and control test}   \label{subsec:systematics}
\begin{figure}
\centering
\includegraphics[scale=0.41]{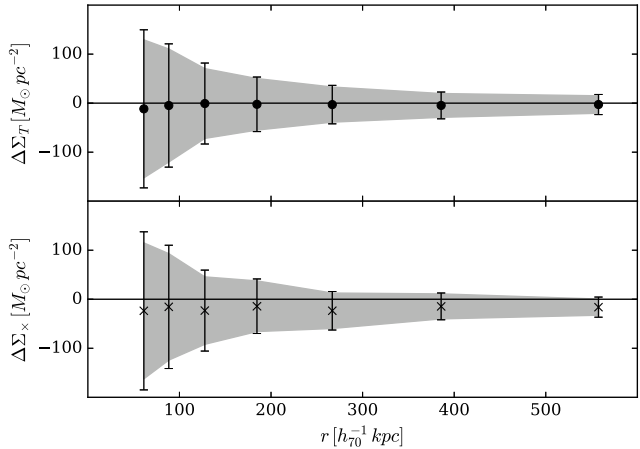}
\caption{Averaged profiles obtained from 200 realisations using random centres for each lensing system. Upper and lower panels show profiles computed by averaging the tangential and cross ellipticity components. The shaded regions correspond to $1\sigma$ dispersion.}
\label{fig:random}
\end{figure}

Here we present the results of a control test to check the confidence of our lensing analysis. We also discuss the uncertainties regarding redshift estimation of background galaxies and the dispersion among stacked groups. We do not take into account errors regarding background sky obscuration given that this effect is negligible for SDSS \citep{Simet2014}. The effects of miscentring are discussed in detail in section \ref{sec:results}.

In order to test the reliability of our measured lensing signal, we compute radial profiles using the background galaxy catalogue centred at random positions within the field of each frame. We carried out 200 relisations to look for any systematics in the density contrast profiles. In Figure \ref{fig:random} we show the averaged profiles together with the dispersion of the resulting 200 relisations. The obtained profiles, using the tangential and cross ellipticity components, are both consistent with a null signal.

Given that the geometrical factor was estimated using a catalogue of photometric redshifts, based on Deep Field 1 which covers 1 square degree,  we estimate the impact of cosmic variance on $\langle \beta \rangle$. We divided this field in 25 non-overlapping areas of $\sim 144$ arcmin$^2$, assuming the average CG redshift of $0.12$, and computed $\langle \beta \rangle$ for each area. The uncertainty in this parameter was estimated according to the dispersion of the 25 regions, obtaining a typical value of $10\%$, which implies a $15\%$ error in the mass.

In order to test the stability of our results we performed a bootstrap analysis by fitting both, SIS and NFW centred models, to 1000 samples of identical size randomly selected with reposition. The distributions of the best fit parameters, $\sigma_{V}$ and $R_{200}$, follow approximately gaussian distributions with dispersions lower than 10\%.

The uncertainties introduced by the issues discussed here are considerably lower than the errors obtained according to the $\chi^2$ dispersion. Nevertheless, these were considered in the final error estimation.

%====================================================================================
\section{RESULTS}		\label{sec:results}
%Results obtained...

%###########################################################################
%###################################
\begin{table*}
\caption{Compact Groups results}\label{tab:esp}
\label{table:1}
\begin{tabular}{@{}ccccccccccccc@{}}
\hline
\hline
\rule{0pt}{1.05em}%
Centre & \multicolumn{2}{c}{SIS}  &   \multicolumn{2}{c}{SIS$_{s}$} & \multicolumn{3}{c}{NFW} & \multicolumn{3}{c}{NFW$_{s}$} &  $S/N$ \\
  
 & $\sigma_{V}$ & $M_{200}$ & $\sigma_{V}$ & $M_{200}$ & $c_{200}$ & $R_{200}$ & $M_{200}$ & $c_{200}$ & $R_{200}$ & $M_{200}$ & \\
% & [$\kms$]  & [$10^{12} h_{70}^{-1} \msun $] & [$\kms$]  & [$10^{12} h_{70}^{-1} \msun $] & [$h_{70}^{-1}$\,Mpc] & [$10^{12} h_{70}^{-1} \msun $] & [$h_{70}^{-1}$\,Mpc] & [$10^{12} h_{70}^{-1} \msun$] & \\

 \hline
\rule{0pt}{1.05em}%
BC & $250\pm60$ & $14\pm8$ & $290\pm60$ & $21\pm13$ & $4.54\pm0.23$ & $0.50\pm0.11$ & $16\pm11$ & $4.38\pm0.22$ & $0.54\pm0.12$ & $21\pm13$ & $4.0$ \\[0.3em] 
GC & $260\pm50$ & $15\pm10$ & $290\pm60$ & $21\pm13$ & $4.51\pm0.22$ & $0.51\pm0.11$ & $17\pm11$ & $4.38\pm0.22$ & $0.54\pm0.12$ & $21\pm14$ & $4.2$ \\[0.3em]
LC & $270\pm40$ & $17\pm8$ & $300\pm60$ & $24\pm12$ & $4.45\pm0.21$ & $0.53\pm0.10$ & $19\pm11$ & $4.33\pm0.21$ & $0.56\pm0.11$ & $22\pm13$ & $4.6$ \\ 

\hline         
\end{tabular}
\medskip
\begin{flushleft}
\textbf{Notes.} Columns: (1) centre choice; (2 - 3) results from the centred SIS fit: velocity dispersion and $M^{SIS}_{200}$; (4 - 5) results from the miscentred SIS fit: velocity dispersion and $M^{SIS}_{200}$; (6 - 8) results from the centred NFW fit: $c_{200}$ estimated with the centred $M^{SIS}_{200}$ (see text for details), $R_{200}$ and $M^{NFW}_{200}$; (9 - 10) results from the miscentred NFW fit: $c_{200}$ estimated with the miscentred $M^{SIS}_{200}$ (see text for details), $R_{200}$ and $M^{NFW}_{200}$; (11) $S/N$ ratio as defined in equation \ref{SN}. $\sigma_{V}$, $R_{200}$ and $M_{200}$ are in units of $\kms$, $h_{70}^{-1}$ Mpc and $10^{12}\,h_{70}^{-1} \msun$, respectively.
\end{flushleft}
\end{table*}
%###################################
%###########################################################################

\subsection{Centre definition analysis}

%###########################################################################
%##################################
\begin{figure}
\centering
\includegraphics[scale=0.41]{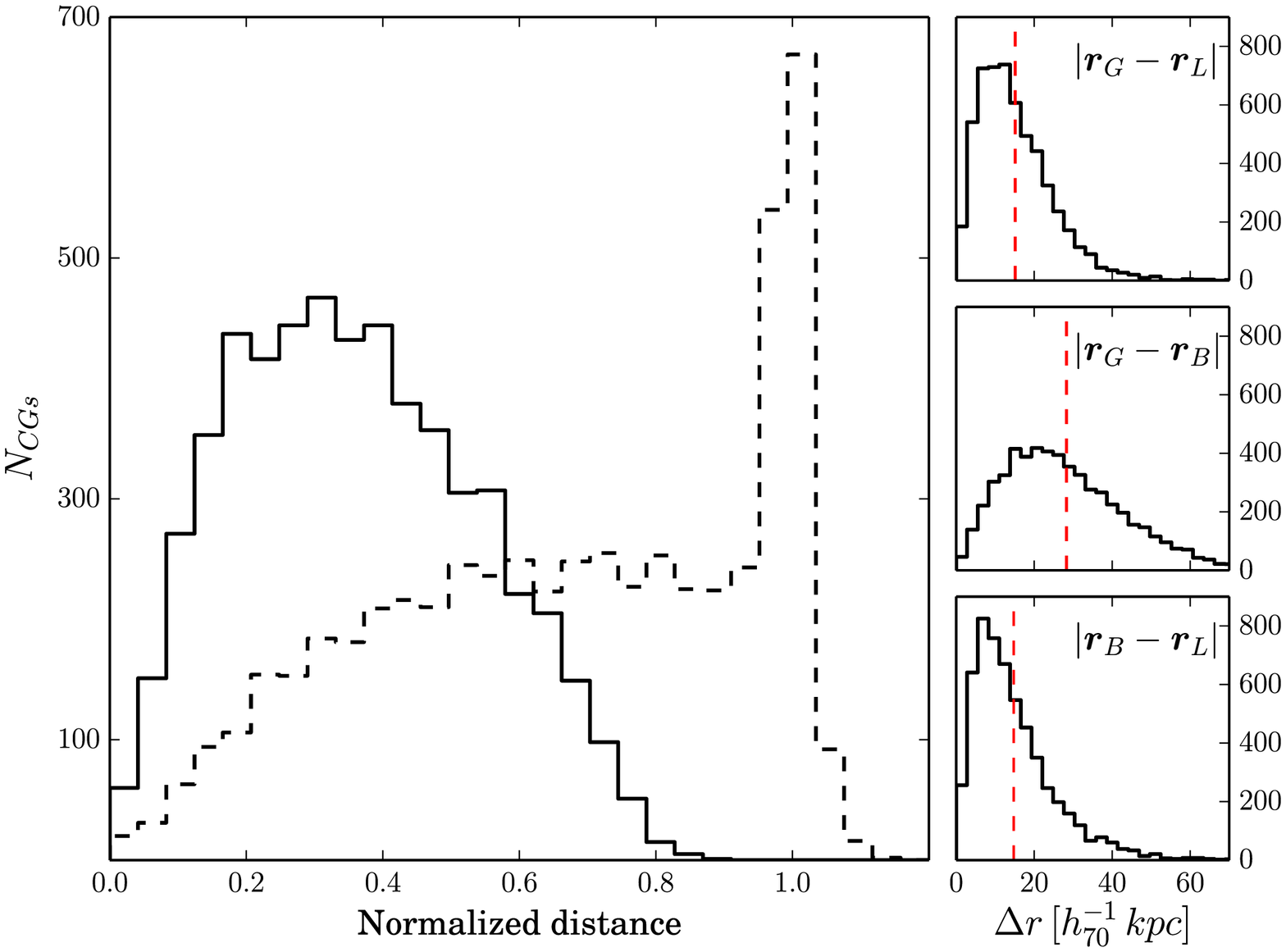}
\caption{\textit{Left}: Distributions of normalized distances. The solid line corresponds to $|\textbf{\textit{r}}_G - \textbf{\textit{r}}_L|/R$, where $\textbf{\textit{r}}_G$ and $\textbf{\textit{r}}_L$ are the coordinates of the geometrical and luminosity weighted centres, respectively, and $R$ is the CG radius. The dashed line corresponds to $|\textbf{\textit{r}}_G - \textbf{\textit{r}}_B|/R$, where $\textbf{\textit{r}}_B$ is the coordinates of the brightest galaxy member. \textit{Right}: distribution of centre differences in physical units. From top to bottom, $|\textbf{\textit{r}}_G - \textbf{\textit{r}}_L|$,  $|\textbf{\textit{r}}_G - \textbf{\textit{r}}_B|$ and $|\textbf{\textit{r}}_B - \textbf{\textit{r}}_L|$. The vertical dashed lines indicate the respective mean values of the distributions.}
\label{fig:centres}
\end{figure}
%##################################
%###########################################################################

%###########################################################################
%###################################
\begin{figure*}
\centering
\includegraphics[width=\textwidth]{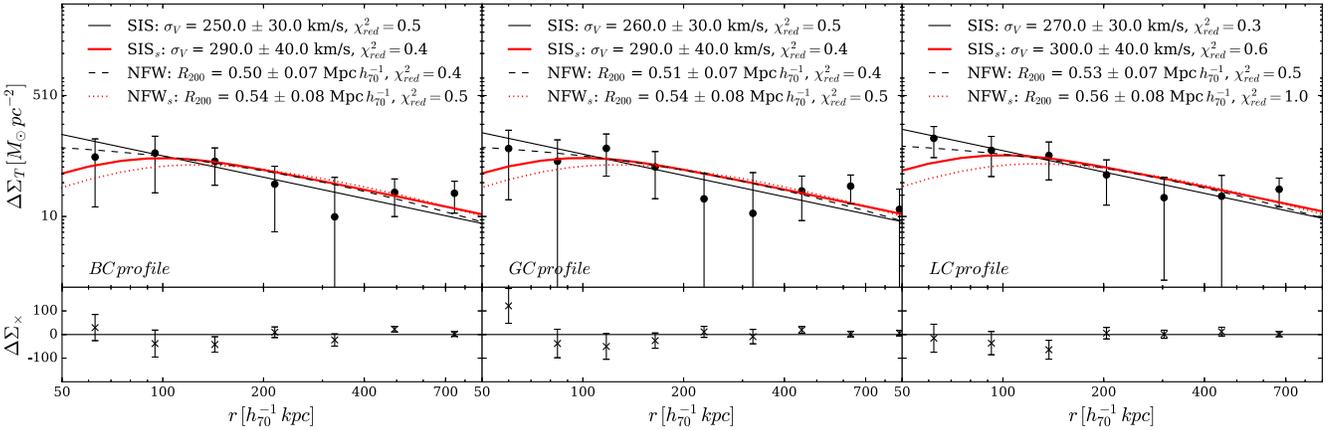}
\caption{Average density contrast $\Delta \Sigma (r)$ profile of CGs sample for each centre: BC (\textit{left}), GC (\textit{middle}) and LC (\textit{right}). Solid thin and thick lines represent the best centred and miscentred SIS fits, respectively; dashed and dotted lines represent the best centred and miscentred NFW fits, respectively. The lower panels of each plot show the profile obtained using the cross component of the background galaxies’ ellipticity. Error bars are computed according to equation \ref{eq:err}. Parameter errors consider only the fitting uncertainties and do not include those discussed in subsection \ref{subsec:systematics}.}
\label{fig:profiles}
\end{figure*}

%###################################
%###########################################################################

In order to analyse the centre offsets with respect to those of the true dark matter halos we consider three different centre choices: the geometrical (GC, included in Catalogue B), the coordinates of the brightest member (BC, also in Catalogue B) and a geometrical centre weighted by luminosity (LC), i.e.:

\begin{equation}
\label{eq:wAverage}
\textbf{\textit{r}}_{L} = \dfrac{\sum \textbf{\textit{r}}_{i}L_{i}}{\sum L_{i}}
\end{equation}
where $\textbf{\textit{r}}_{i} = (\alpha, \delta)$ are the group members celestial coordinates and $L_{i}$ are their corresponding $r$-band luminosities. $L_{i}$ were computed using CGs' redshifts and $r$-band magnitudes corrected by galactic extinction. We applied k-corrections to magnitudes, using \citet{Chilingarian2010} public code \texttt{calck\_cor.py}\footnote{Avialble at: http://kcor.sai.msu.ru/getthecode/}. In Figure \ref{fig:centres} we show the distributions of normalized centre differences and in physical units: $|\textbf{\textit{r}}_G - \textbf{\textit{r}}_L|$ (where $\textbf{\textit{r}}_G$ is the coordinates of the geometrical centre); $|\textbf{\textit{r}}_G - \textbf{\textit{r}}_B|$ (where $\textbf{\textit{r}}_B$ is the coordinates of the brightest galaxy member) and $|\textbf{\textit{r}}_B - \textbf{\textit{r}}_L|$. As can be noticed, the distribution of the brightest galaxy shows a peak at the group radius given the characteristics of the identification algorithm of CGs.

%The tangential ellipticity components and distance of the background galaxies were calculated considering the celestial coordinates of the GC, BC and LC centres.
% and 

The measured density profiles for the three centre choices are shown in Figure \ref{fig:profiles}. 
We include in this Figure the fitted centred (SIS and NFW) and miscentred (SIS$_s$ and NFW$_s$) models, with their corresponding parameters and the reduced $\chi^{2}$ values of each fit. Points and crosses represent the tangential and cross density contrast components averaged in annular bins, respectively.

As it can be seen, there are differences in the inner region of the derived profiles. The slope of the LC centred profile presents no signs of flattening inwards ($r\lesssim 100 \, h^{-1}_{70} \kpc$), contrary to GC and BC centred profiles. Nevertheless, according to $\chi_{red}^2$, both profiles are well described by a miscentred model as well as by a centred one. In general, derived masses from both centred and miscentred profiles are in mutual agreement taking into account the uncertainties, while larger differences are observed for SIS masses. Given that the SIS profile is more sensitive to centre definition, we have compared the obtained $\chi^2$ of both, centred and miscentred, SIS fitted profiles (see Figure \ref{fig:profiles}), and therefore we choose the LC as the gravitational potential centre. In Table\,\ref{table:1} we summarise our results adding the errors discussed in subsection \ref{subsec:systematics}.

The model that best describes the LC centred profile is the centred SIS yielding an average velocity dispersion of $\sigma_{V} = 270 \pm40 \kms$, which corresponds to $M_{200} = 17 \pm 8 \times 10^{12} \,h_{70}^{-1} \msun$. Since the halos of CGs are expected to have undergone significant contraction due to the baryonic cooling and collapse, a SIS profile can be a suitable alternative model to NFW, to describe the mass distribution of these low mass systems. It should be noted, however, that the estimated SIS and NFW masses are in good agreement within a $\sim 10 \%$ factor as in previous works \citep{Eli2015, Eli2016}. For the rest of the analysis we use these fitted parameters to compare them with dynamical estimates and to study variations in the total sample.

%###########################################################################
%###################################
\begin{figure}
\centering
\includegraphics[scale=0.42]{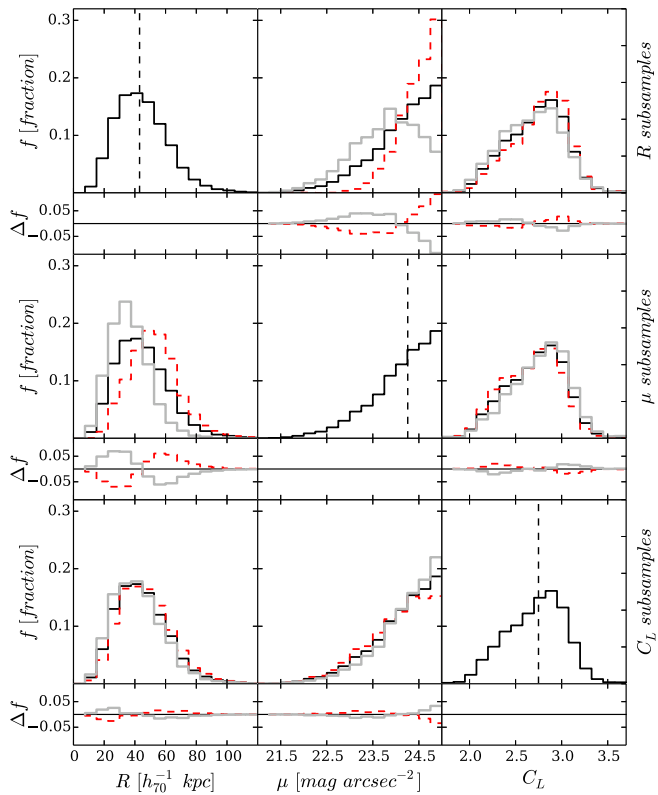}
\caption{Parameter variation for each subsample. Columns: (1) physical radius, $R$; (2) surface brightness, $\mu$; (3) weighted concentration index, $C_L$ (see text for definition). Rows: (1) $R$ subsamples; (2) $\mu$ subsamples; (3) $C_L$ subsamples. All distributions were normalized to have the same area. The solid black lines correspond to the complete sample; the dashed and gray lines correspond to the higher and lower subsamples, respectively. Below each panel, we show the residuals between the complete sample distribution and each subsample.}
\label{fig:distributions}
\end{figure}
%###################################
%###########################################################################

\subsection{Dependence of the lensing signal on CGs physical properties}
%###########################################################################
%###################################
\begin{table}
\caption{Subsample results.}
\label{tab:subsamples}
\begin{small}
\begin{center}
\begin{tabular}{@{}cccc@{}}

\hline
\hline
\rule{0pt}{1.05em}%
Subsample & SIS  &  \multicolumn{2}{c}{NFW} \\
  
 & $\sigma_{V}$ & $c_{200}$ & $R_{200}$  \\
 
\hline
\rule{0pt}{1.05em}%
$R > 43 \; h_{70}^{-1}$ kpc & $270\pm50$  & $4.44\pm0.24$ & $0.49\pm0.13$  \\[0.3em] 
$R < 43 \; h_{70}^{-1}$ kpc & $260\pm60$  & $4.50\pm0.25$ & $0.51\pm0.13$  \\[0.8em]
%\hline
$\mu > 24.25 \magsec$ & $290\pm50$  & $4.37\pm0.22$ & $0.58\pm0.13$  \\[0.3em]
$\mu < 24.25 \magsec$ & $240\pm60$  & $4.59\pm0.27$ & $0.45\pm0.13$  \\[0.8em] 
%\hline
$C_L > 2.75$ & $300\pm50$ & $4.33\pm0.21$ & $0.56\pm0.13$  \\[0.3em] 
$C_L < 2.75$ & $220\pm60$ & $4.70\pm0.31$ & $0.43\pm0.14$  \\
\hline     
\end{tabular}
\end{center}
\begin{flushleft}
\textbf{Notes.} Columns: (1) selection criterium according to the median value of each distribution; (2) velocity dispersion from the SIS fit $[\kms]$; (3 - 4) fixed $c_{200}$ and estimated $R_{200}$ $[h_{70}^{-1}$ Mpc] from the NFW fit.
\end{flushleft}
\end{small}
\end{table}
%###################################
%###########################################################################
We studied how CGs average lensing mass varies with respect to three parameters: physical radius, $R$, surface brightness, $\mu$, and average concentration index weighted by luminosity, $C_{L}$. We defined $C_{L}$ for a group as:
\begin{equation}
C_L = \dfrac{\sum c_i L_i}{\sum L_i}
\end{equation}
where $c_i$ is the individual concentration index of member galaxies defined as the ratio of the radii enclosing 90\% and 50\% of the Petrosian flux, i.e. $c_i = r_{90}/r_{50}$. For each parameter we divided our sample into two equal-sized subsamples according to the median value of the parameter distribution.  In Figure \ref{fig:distributions} we plot these parameters distributions together with their respective subsamples distributions. 

In Table \ref{tab:subsamples} we summarize the results of this analysis. To test the significance of these results, we performed a jacknife resampling technique by randomly choosing 1000 subsamples taking 50\% of the groups. From this analysis we obtained gaussians distributions for the fitted parameters with dispersions of $30 \kms$ and $0.08 \mpc$ for $\sigma_{V}$ and $R_{200}$, respectively. We find no significant variation of the fitted parameters for the $R$ and $\mu$ subsamples, since they are in good agreement taking into account the errors. However, for the $C_{L}$ subsamples, the resulting parameters differ by $\sim 2\sigma$ considering the jacknife dispersion.

\begin{figure}
\centering
\includegraphics[scale=0.41]{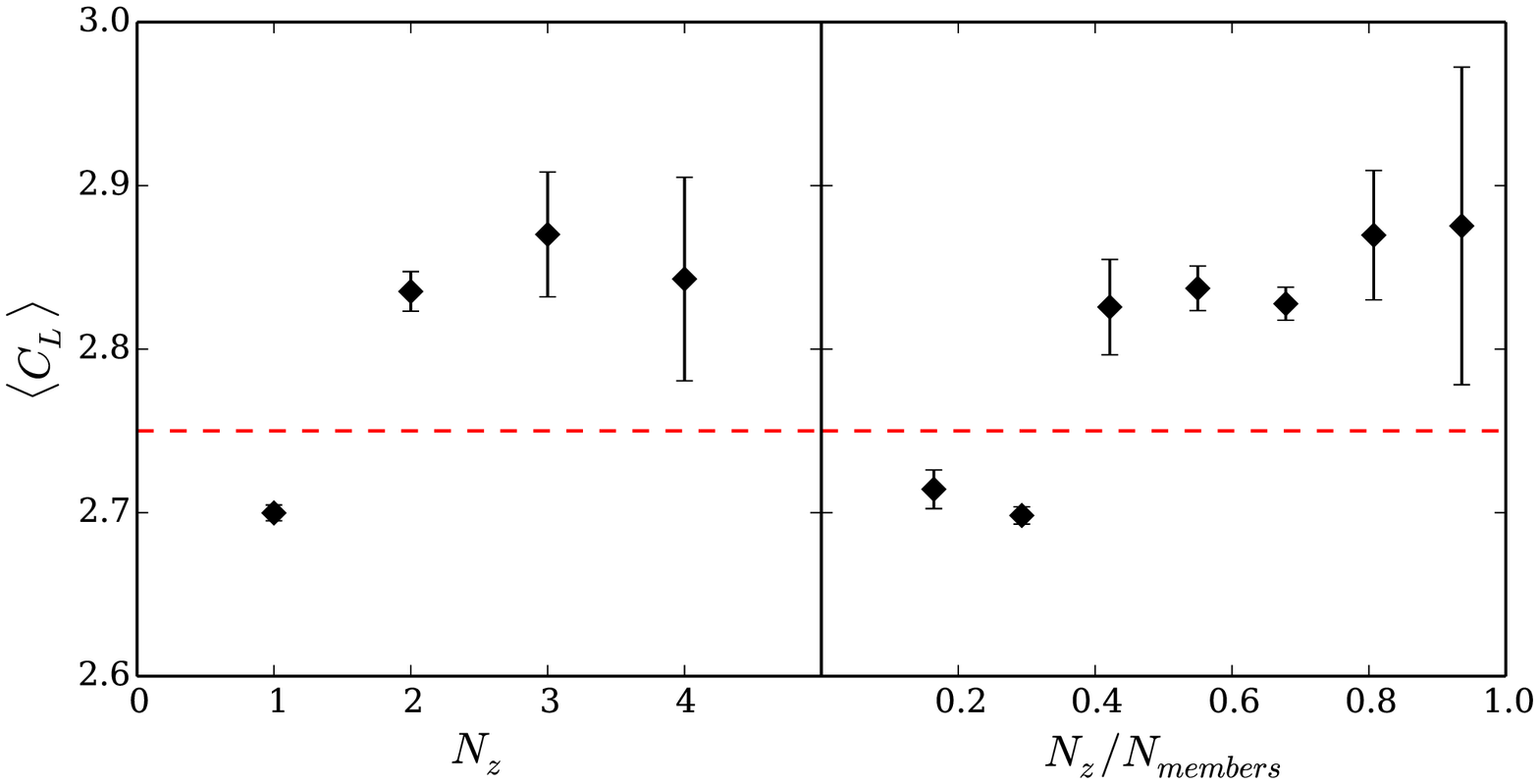}
\caption{Variation of the average $C_L$ with the number of members with spectroscopic redshift (\textit{left}) and with the respective fraction (\textit{right}). The dashed line marks the median value of $C_L$ used to divide the sample.}
\label{fig:C_N}
\end{figure}

\begin{figure*}
\begin{multicols}{2}
\centering
\def \scaleGX {0.3}
\includegraphics[scale=\scaleGX]{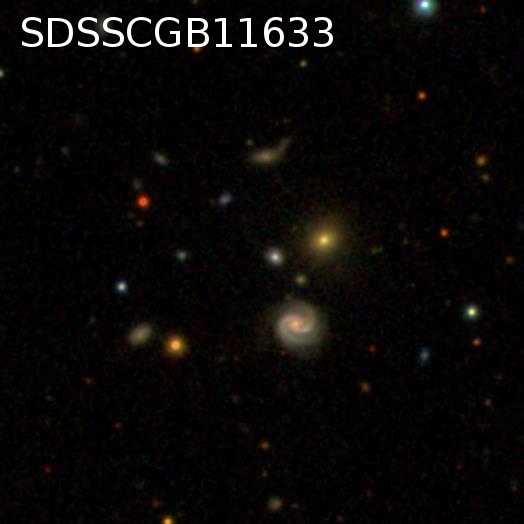}
\includegraphics[scale=\scaleGX]{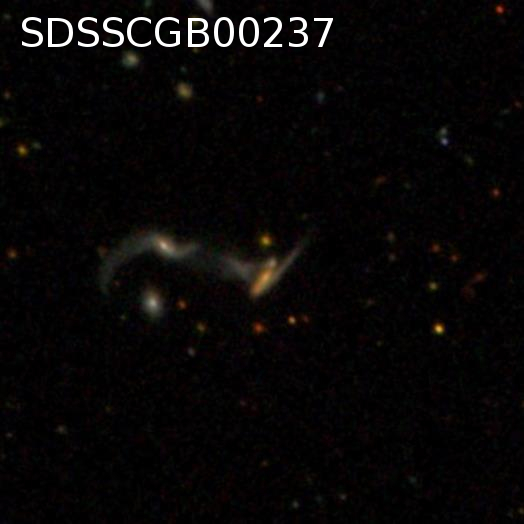}\\ \,
\includegraphics[scale=\scaleGX]{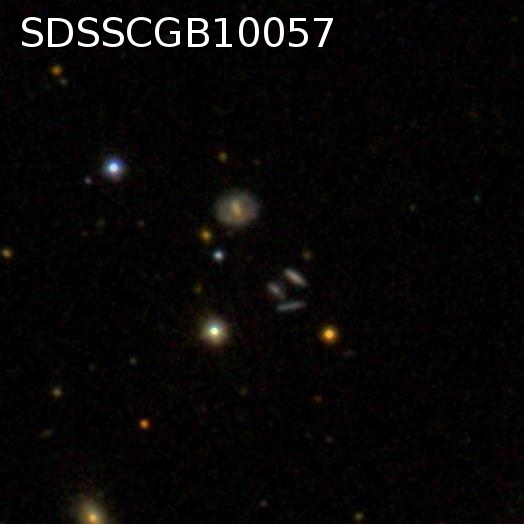}
\includegraphics[scale=\scaleGX]{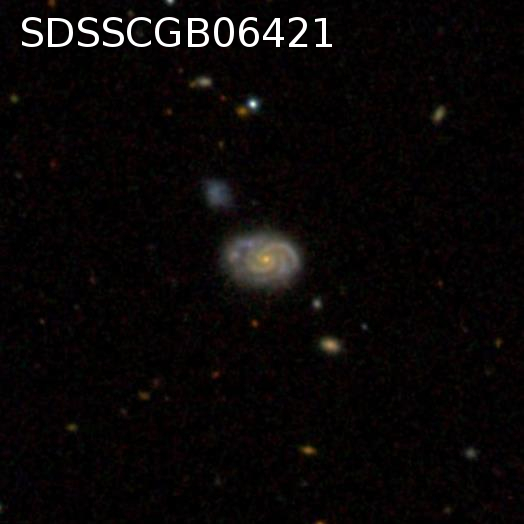}
\includegraphics[scale=0.4]{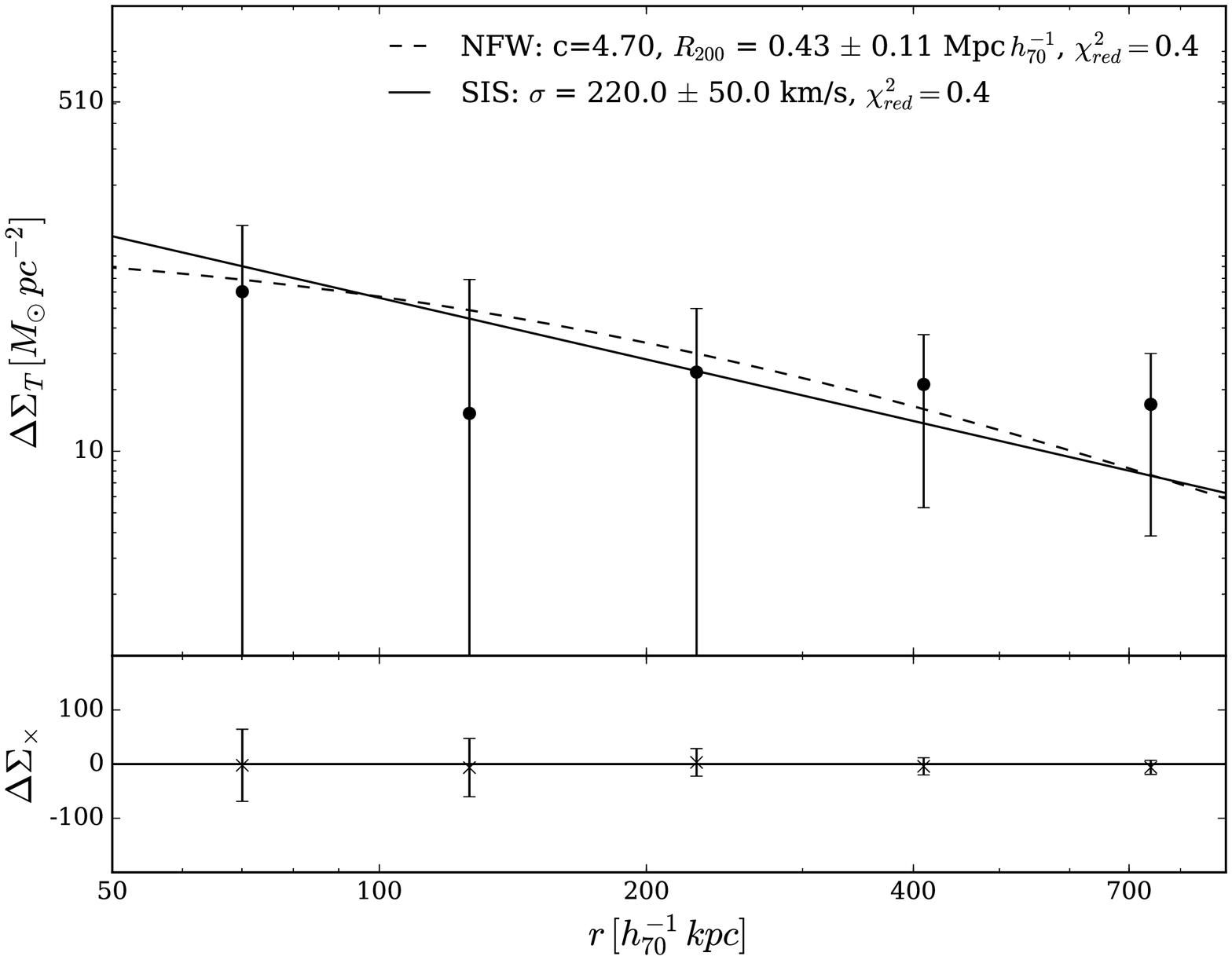}

\includegraphics[scale=\scaleGX]{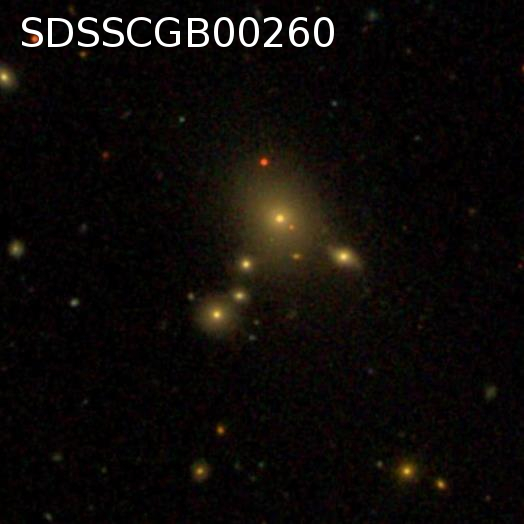}
\includegraphics[scale=\scaleGX]{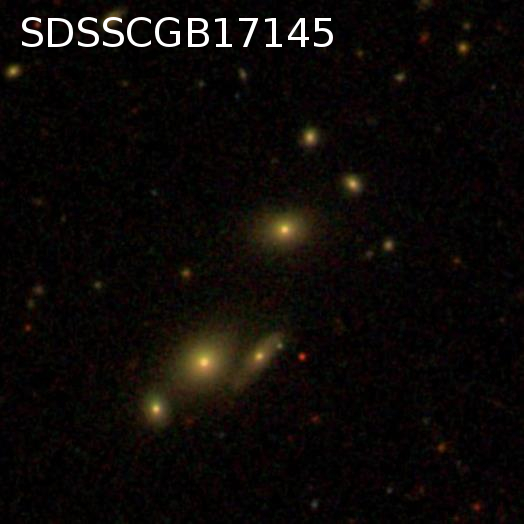}\\ \,
\includegraphics[scale=\scaleGX]{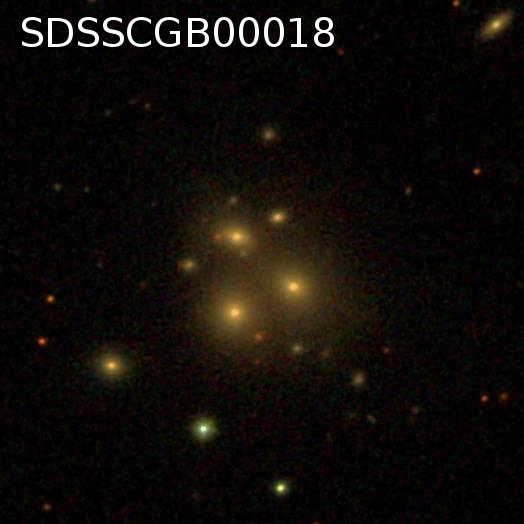}
\includegraphics[scale=\scaleGX]{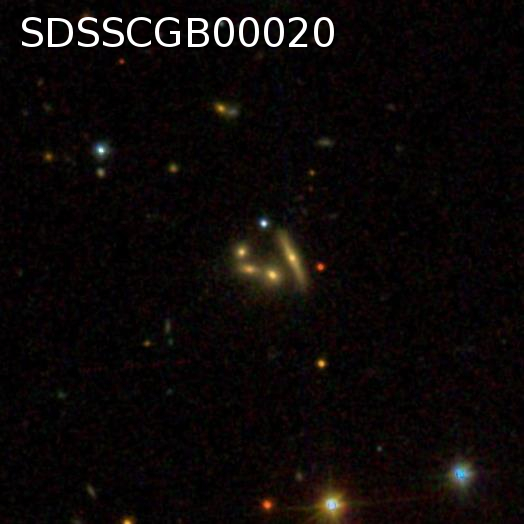}
\includegraphics[scale=0.4]{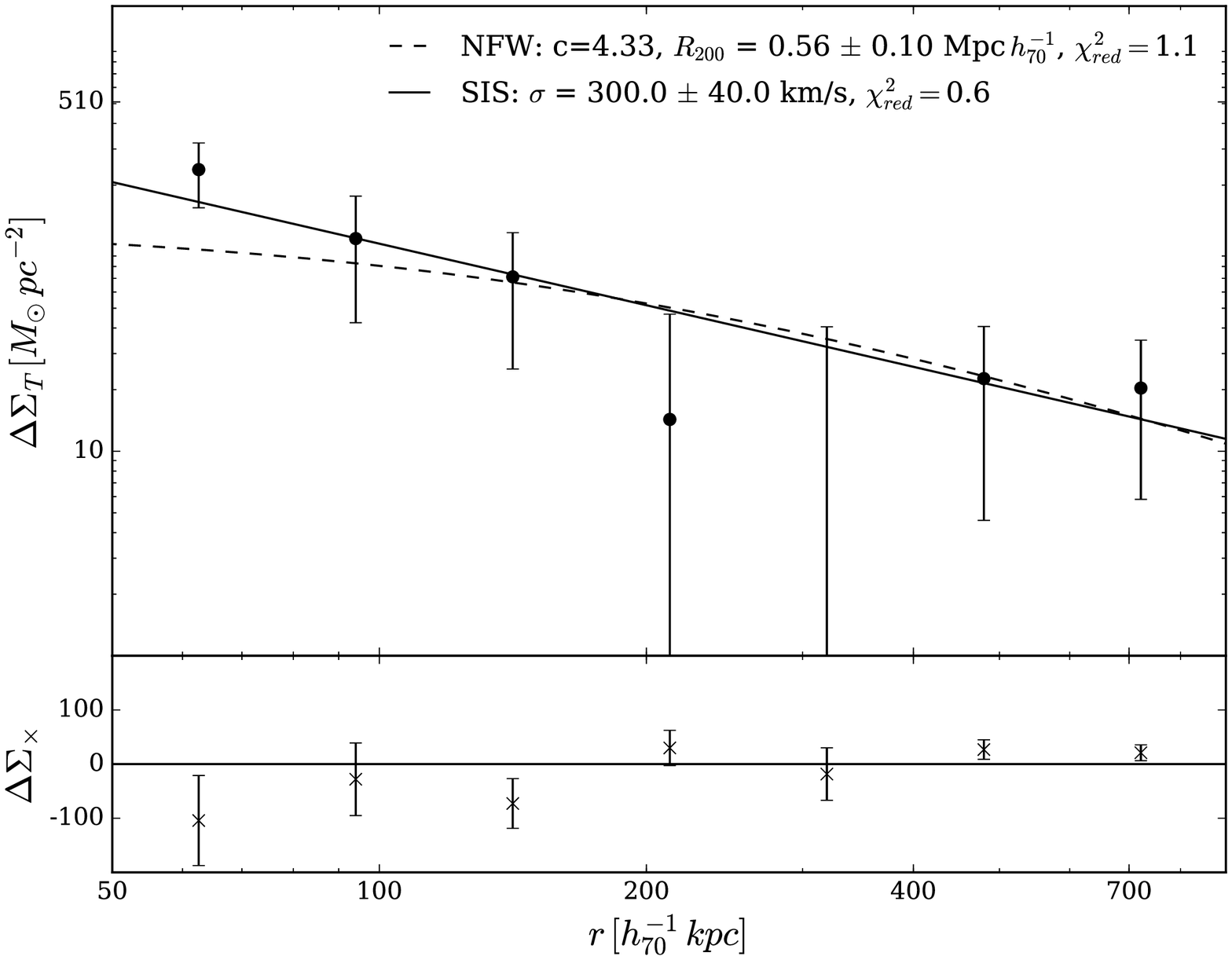}
\end{multicols}
\caption{Example of CGs present in both $C_L$ subsamples accompanied by their respective density contrast profile. As in Figure \ref{fig:profiles}, parameter errors consider only the fitting uncertainties and do not include those discussed in subsection \ref{subsec:systematics}. The four images on the left, and the profile below them, correspond to the sample with lower $C_L$ values. The remaining figures on the right side correspond to systems with higher $C_L$ values. Images were obtained from the SDSS Navigate Tool.}
\label{fig:profiles_CL}
\end{figure*}

The concentration index is an indicator of galaxy morphology, where late-type galaxies tend to have lower $c_i$ values than early-type. Thus, groups with lower and higher $C_{L}$ are expected to be dominated by late and early type galaxies, respectively. The detection of a higher lensing signal for groups with higher $C_{L}$ values could be influenced by a lower fraction of interlopers. Given that CGs are expected to have a greater fraction of early type members, by selecting CGs with low $C_{L}$ we could be including more systems with interlopers and, thus, reducing the lensing signal. As a matter of fact, this cut in concentration modifies the distribution of surface brightness: higher $C_L$ groups tend to be brighter than lower $C_L$ groups (see Figure \ref{fig:distributions}). As mentioned before, the fraction of interlopers declines as brighter groups are considered \citep{McConnachie2008}, and since the estimated parameters are in agreement for both $\mu$ subsamples, this result suggests that a cut in $C_{L}$ may be more efficient than a cut in $\mu$ in order to reduce the contamination in the CGs sample. This is also evident from the observed relations between $\langle C_L \rangle$ vs. $N_z$, and $\langle C_L \rangle$ vs. $N_z/N_{members}$. $N_z$ is the number of member galaxies with available spectroscopy, and $N_{members}$ is the total number of members \citep[we restrict to groups with a maximum line-of-sight velocity difference between pairs of members, $max(\Delta v) < 1000 \kms$, a usual criterium to minimize interlopers ][]{Hickson1992,McConnachie2009}. As can be seen in Figure \ref{fig:C_N}, groups with higher $C_L$  tend to have higher $N_z$ and $N_z/N_{members}$ values, making them more reliable. In Figure \ref{fig:profiles_CL} we show images for both subsamples together with  their respective  average density constrast profiles. By selecting CGs dominated by early-type galaxies, the systems tend to be more masive and evidence a more evolved structure. 
 
\subsection{Comparison with dynamical estimates}
Given that the $\sigma_V$ parameter derived from the weak lensing analysis can be directly compared with dynamical estimates, we have analysed the redshift distribution of CGs' member galaxies in order to estimate the dynamical velocity dispersion, $\sigma_{V,dyn}$. With this aim, we consider only CGs having 3 or more members with redshift information and, as before, we discard those with $max(\Delta v) > 1000 \kms$. From our sample of 5568 CGs, only 61 satisfy these requirements. We find a median dynamical velocity dispersion  $\sigma_{V,dyn} = 224 \pm 13 \kms$, where the uncertainty corresponds to the $1\sigma$ standard deviation derived from 1000 bootstrap resamplings. This value is in good agreement with other dynamical estimates for CGs: $\simeq 200 \kms$ \citep{Hickson1992,Duplancic2015,Sohn2015} and $\simeq 230 \kms$ \citep{McConnachie2009}.

Since gravitational lensing allows the measurement of the mass distribution at large angular distances from the centre, one would expect that CGs lensing inferred velocity dispersions could be higher than those derived from their core's dynamics. It should also be taken into account that the presence of dynamical friction among highly interacting group members could further reduce their velocity dispersion. Nevertheless, the weak lensing estimate of $\sigma_{V} = 270 \pm 40 \kms$, although slightly higher, mutually agree with dynamical determinations within $1\sigma$.

Using the same criteria, we also estimated the dynamical velocity dispersion for both $C_L$ subsamples. For groups with higher $C_L$ values we find $\sigma_{V,dyn} = 238 \pm 15 \kms$, while for groups with lower $C_L$ we find $\sigma_{V,dyn} = 190 \pm 22 \kms$. These results show the same tendency as the aforementioned weak lensing estimates, reinforcing their interpretation.

%====================================================================================
\section{SUMMARY}			
\label{sec:discussion}

In this work we analysed a sample of Compact Groups from \citet{McConnachie2009} Catalogue B using weak lensing stacking techniques. We derive the average density contrast profile of the composite system for three centre definitions: the geometrical centre, the brightest galaxy member and a luminosity weighted centre. Measured profiles were fitted using centred and miscentred, SIS and NFW, density models. Luminosity weighted centres were selected as the best description of the true dark matter halo centres.

We also studied the lensing signal dependence on physical parameters (radius, surface brightness and concentration index of galaxy members) of the CGs. We did not observe a significant difference between the fitted parameters for subsamples defined according to group radius and surface brightness cuts. Nevertheless, CGs composed by galaxies with larger $c_i$ show a stronger lensing signal. This could be explained by a lower number of interlopers, as well as by a trend to include more massive and evolved systems. We argue that by considering groups with higher concentration index weighted by luminosity, could be efficient in order to increase the fraction of genuine CGs in the sample.

The resulting velocity dispersion derived from the SIS profile was compared to the dynamical estimate obtained from spectroscopic information of member galaxies. Although the lensing estimate is slightly higher, both results are in good agreement within uncertainties.

%In order to perform an independent analysis we studied a second sample of more distant of CGs. We use the Canada-France-Hawaii Telescope Legacy Survey (CFHTLS), where we can rely on its deeper expositions to increase the density of background galaxies and thus, the lensing signal. We cross-correlated the CFHTLS wide-field database with an extended sample of CGs up to $z = 0.5$ and with no surface brightness restriction. CFHTLS covers only 150 square degrees of sky, roughly $\sim 1\%$ of the sky covered by SDSS-DR6, considerably reducing the number of imaged CGs to 25. Despite deeper and better quality images, the stacked density of background galaxies was insufficient to boost the signal-to-noise ratio above the detectable threshold. The main limitation is the low number of CGs available in CFHTLS data. We estimate that in order to obtain a S/N comparable with the value measured from the SDSS sample, it would be necessary to stack $\sim 400$ CGs, more than an order of magnitude than public CFHTLS data.

This work provides the first lensing analysis of a sample of CGs based on SDSS images. Our results, in agreement with other dynamical estimates, give hints on the mass distribution and dependence on CGs properties. In a forthcoming paper we will consider in detail mass-to-light ratio and a comparison to simulations.

%====================================================================================
\section*{Acknowledgments}

We thank the anonymous referee for the very useful comments that improved the content and clarity of the manuscript. This work was partially supported by the Consejo Nacional de Investigaciones Cient\'{\i}ficas y T\'ecnicas (CONICET, Argentina) 
and the Secretar\'{\i}a de Ciencia y Tecnolog\'{\i}a de la Universidad Nacional de C\'ordoba (SeCyT-UNC, Argentina).

Funding for SDSS-III has been provided by the Alfred P. Sloan Foundation, the Participating Institutions, the National Science Foundation, and the U.S. Department of Energy Office of Science. The SDSS-III web site is http://www.sdss3.org/.

SDSS-III is managed by the Astrophysical Research Consortium for the Participating Institutions of the SDSS-III Collaboration including the University of Arizona, the Brazilian Participation Group, Brookhaven National Laboratory, Carnegie Mellon University, University of Florida, the French Participation Group, the German Participation Group, Harvard University, the Instituto de Astrofisica de Canarias, the Michigan State/Notre Dame/JINA Participation Group, Johns Hopkins University, Lawrence Berkeley National Laboratory, Max Planck Institute for Astrophysics, Max Planck Institute for Extraterrestrial Physics, New Mexico State University, New York University, Ohio State University, Pennsylvania State University, University of Portsmouth, Princeton University, the Spanish Participation Group, University of Tokyo, University of Utah, Vanderbilt University, University of Virginia, University of Washington, and Yale University.

This research has made use of NASA's Astrophysics Data System and Cornell University arXiv repository.

We made an extensive use of the following python libraries:  http://www.numpy.org/, http://www.scipy.org/, http://roban.github.com/CosmoloPy/ and http://www.matplotlib.org/.\\

\bibliographystyle{mn2e}
\bibliography{references}

\appendix

\end{document}